
%
%
%
%
%
\input amsppt.sty


%
%
%
%
%
%
%

\def\qed{\hbox{${\vcenter{\vbox{
    \hrule height 0.4pt\hbox{\vrule width 0.4pt height 6pt
    \kern5pt\vrule width 0.4pt}\hrule height 0.4pt}}}$}}

\def\and{\  \hbox{ and } \ }

\def\la{\lambda}

\def\al{\alpha}

\def\De{\Delta}

\def\0{\bold{0}}




\title Structure Constants
\endtitle

\topmatter
\title Remarks on
the structure constants of the Verlinde algebra associated to $sl_3$
\endtitle

\author G.Felder and A.Varchenko
\endauthor
\date February, 1995 \enddate
\thanks Supported in part by NSF grants DMS-9400841 and DMS-9203929
\newline
Department of Mathematics, University of North Carolina, Chapel Hill,
N.C. 27599, \newline
felder\@math.unc.edu, varchenko\@math.unc.edu.
\endthanks
\endtopmatter

\document

The Verlinde fusion algebra is an associative commutative algebra associated
to a Wess-Zumino-Witten model of conformal field theory [V,F,GW,K,S].
Such a model is
labelled by a simple Lie algebra
$\frak g$ and a natural number $k$ called level. The Verlinde algebra
$A(\frak g,k)$ is a finitely generated algebra
with generators $V_{\lambda}$ enumerated
by irreducible $\frak g$-modules admissible for
the model. The structure constants $N_{\lambda, \mu}^{\nu}$ of the
multiplication
$V_{\lambda} \cdot V_{\mu} = \sum_{\nu}  N_{\lambda, \mu}^{\nu} V_{\nu}$
are non-negative integers important for applications.
 ( We use the formula in [K, Sec.13.35] as a
definition of the structure constants.)

 \subhead Example 1 \endsubhead
The algebra $A(sl_2,k)$ has $k+1$ generators
$V_0,...,V_k$. For fixed $\lambda, \nu$ and varying $\mu$,
the structure constants  $N_{\lambda, \mu}^{\mu+\nu}$ are either zero or form
the characteristic function of an interval with respect to $\mu$.
Namely,  $N_{\lambda, \mu}^{\mu+\nu}=0$, if $\lambda-\nu$ is odd or if
$|\nu|>\lambda$. If $\lambda-\nu$ is even and $|\nu|<\lambda$, then
 $N_{\lambda, \mu}^{\mu+\nu}=1$ for $\mu \in [(\lambda-\nu)/2,
k-(\lambda+\nu)/2]$ and  $N_{\lambda, \mu}^{\mu+\nu}=0$ otherwise.

It is interesting that after an affine change of the variable the function
 $N_{\lambda, \mu}^{\mu+\nu}$ of $\mu$ becomes the weight function
of the irreducible $sl_2$-module with highest weight $k-\la$.

In this paper we give a similar formula for the
structure constants of the Verlinde algebra associated to $sl_3$.

\subhead 1. Weight Functions
\endsubhead

Let $\Cal P= \Bbb Z^3/\Bbb Z \cdot (1,1,1)$
be the two dimensional weight lattice of $sl_3$. Let $L_1=(1,0,0), L_2=(0,1,0),
L_3=(0,0,1), \alpha_1=(1,-1,0), \alpha_2=(0,1,-1), \alpha_3=(-1,0,1)$,
 considered
as elements of $\Cal P$.

For a natural number $k$ introduce coordinates on $\Cal P$:
$$ y_1(\mu):=(\alpha_1,\mu), \qquad y_2(\mu):=(\alpha_2,\mu), \qquad
y_3(\mu):=k+(\alpha_3,\mu)=k-y_1-y_2, $$
where $(x,y)=x_1y_1+x_2y_2+x_3y_3$.

 \subhead Definition 1 \endsubhead
{\it  A triangle} in $\Cal P$ is a set $\Delta$ of the form
$$ \Delta = \{ \mu \in \Cal P \, |\, y_i(\mu) \geqslant  A_i, i=1,2,3\} $$
for some integers $A_i.$

The number $k-A_1-A_2-A_3$ is called {\it the size} of the triangle. It is the
integral length of its edges.

 \subhead Definition 2 \endsubhead
A pair consisting of a natural number $m$ and a triangle
$\Delta$ of size $\ell$ is called {\it appropriate} if
$\ell \geqslant 2m-2$.

 \subhead Definition 3 \endsubhead
{\it The weight function}  $w_{m,\Delta}$ associated to
an appropriate pair $m, \Delta$ is the following function
$$ w_{m,\Delta} : \Cal P \to \Bbb Z_{\geqslant 0}, $$
which is zero outside $\Delta$, and its level sets
inside $\Delta$ are shown in the picture. The level sets of
$w_{m,\Delta}$ are defined inductively. The function $w_{m,\Delta}$
is equal to zero at the corner  triangles of $\Delta$ of size $m-2$.
$w_{m,\Delta}$ is equal to 1 at
the boundary integral points of the remaining part
of $\Delta$.
Denote by $n$ the number $ min\{m-1, \ell-2m+2\}$.
Assume that the points of $\Delta$ where $w_{m,\Delta}<j$
for $j < n$ are already defined,  define the set
 $w_{m,\Delta}=j$
as the set of boundary integral points of the convex hull of the
remaining integral points of $\Delta$.
If the set  $w_{m,\Delta}=n$
is already defined,  put  $w_{m,\Delta}=n+1$ at the remaining part of
$\Delta$.

\subhead 2. Main Result \endsubhead

Fix a natural number $k$. A weight $\lambda \in \Cal P$ is called
{\it admissible} at
level $k$ if $y_i(\lambda) \geq 0, i=1,2,3.$
Denote by $V_{\lambda}$ the irreducible $sl_3$-module with highest weight
$\lambda$.
For $\lambda, \nu \in \Cal P$, denote by $d_{\lambda}(\nu)$
the dimension of the weight subspace of $V_{\lambda}$ of weight
$\nu$.

The generators of the Verlinde algebra $A(sl_3,k)$ are labelled by irreducible
$sl_3$-modules $V_{\mu}$ with admissible highest weights.

Let  $N_{\lambda, \mu}^{\nu}$ be the structure constants of $A(sl_3,k)$.
Here  $\lambda, \mu, \nu \in \Cal P$,
and we set  $N_{\lambda, \mu}^{\nu}=0$, if at least one of the indices
is not admissible.

For fixed  $\lambda, \nu $
consider  $N_{\lambda, \mu}^{\mu+\nu}$ as a function of $\mu \in \Cal P$.

\proclaim {Theorem 1}

1. If   $d_{\lambda}(\nu)=0$, then  $N_{\lambda, \mu}^{\mu+\nu}=0.$

2. If   $d_{\lambda}(\nu)>0$, then  there is an appropriate pair
$m, \Delta$ such that $m=d_{\lambda}(\nu)$ and
$$ N_{\lambda, \mu}^{\mu+\nu} = w_{m,\Delta}(\mu)$$
for all $\mu$.

\endproclaim

Below we describe the triangle $ \Delta$.

Assume that  $d_{\lambda}(\nu)>0$.  Set
$$
z_{i,\lambda}(\nu) = min\{ |m|-1 \, | \, m \in \Bbb Z,
d_{\lambda}(\nu + m\alpha_i)<d_{\lambda}(\nu)\}
$$
for $i=1,2,3.$

 \subhead Definition 4
 \endsubhead
 A point $\nu \in \Cal P$ is of {\it type} $I$
(resp. $II$) if the product \newline
$(\alpha_1, \nu)\cdot(\alpha_2, \nu)\cdot(\alpha_3, \nu)$
is non-positive (resp. non-negative).

For $\nu$ of type $I$, let $i$ and $j$ be such that
$(\alpha_i, \nu) \geq 0, (\alpha_j, \nu)
\geq 0$.
For $\nu$ of type $II$, let $i$ and $j$ be such that
$(\alpha_i, \nu) \leq 0, (\alpha_j, \nu)
\leq 0$.
In both cases let $\ell$ be the remaining index in $\{1,2,3\}$.

\proclaim {Theorem 2} Assume that  $d_{\lambda}(\nu)>0$. Then
the triangle $\Delta$ in
Theorem 1 has the following form. If
$\nu$ is of type I, then
$$ \Delta =
\{ \mu \in \Cal P \, | \,
y_i(\mu) \geq z_{i,\lambda}(\nu), \,
y_j(\mu) \geq z_{j,\lambda}(\nu), \,
y_{\ell}(\mu) \geq z_{\ell,\lambda}(\nu) - (\alpha_\ell, \nu) \}.
$$

If $\nu$ is of type II, then
$$
\Delta =
\{ \mu \in \Cal P \, | \,
y_i(\mu) \geq z_{i,\lambda}(\nu) - (\alpha_i, \nu), \,
y_j(\mu) \geq z_{j,\lambda}(\nu) - (\alpha_j, \nu), \,
y_{\ell}(\mu) \geq z_{\ell,\lambda}(\nu) \}.
$$

\endproclaim

\subhead Example 2
\endsubhead
Let $\nu=0$.  Then $d_{\lambda}(0)=0$ unless
$\lambda = (a+3b)L_1-aL_3$ or
$\lambda =aL_1-(a+3b)L_3$ for some non-negative integers $a$ and $b$.
If $\lambda$ has this form, then $d_{\lambda}(0)=a+1$
and the triangle of Theorem 1 is
$$
\Delta =
\{ \mu \in \Cal P \, | \,
y_i(\mu) \geq b, i=1,2,3 \}.
$$

 \subhead Remark  \endsubhead For an irreducible $sl_3$-module $V_{\lambda}$,
 consider its weight function
$d_{\lambda} : \Cal P \to \Bbb Z_{\geq 0}.$
It is easy to see that, after an affine change of variables,
the function $d_{\lambda}$ becomes the weight function
 of  an appropriate pair $m, \Delta$, cf.
 [FH, Sec. 13].
Namely, the affine change of variables $r_{\lambda} :
\Cal P \to \Cal P, \ell_1L_1+\ell_2L_2 \mapsto
(\ell_1+\ell_2)\alpha_1+\ell_2\alpha_2+\lambda,$
transforms $d_{\lambda}$ into the weight function
$w_{m,\Delta}$, where $m=(\alpha_1,\lambda)+1$ and $\Delta$
is a triangle of size $(\alpha_1+2\alpha_2,\lambda)$.

Conversely, any weight function
$w_{m,\Delta}$ after a suitable affine change of variables
becomes the weight function of an irreducible $sl_3$-module.

 \subhead Remark  \endsubhead It would be interesting to find an analog of
these
theorems for the $sl_4$-Verlinde algebra.

\subhead 3. Application \endsubhead

Consider the Wess-Zumino-Witten model associated to $sl_3$ at level $k$.
Consider the space of conformal blocks associated to a torus with one
marked point labelled by an admissible $sl_3$-module $V_{\lambda}.$
Denote by $D(\lambda,k)$ the dimension of this space.
 From the fusion rules [TUY], it follows that
 $D(\lambda,k)= \sum_{\mu}N_{\lambda,\mu}^{\mu}$.

\proclaim {Corollary }  $D(\lambda,k)=0$ unless
$\lambda = (a+3b)L_1-aL_3$ or
$\lambda =aL_1-(a+3b)L_3$ for some non-negative integers $a$ and $b$.
If $\lambda$ has this form, then
$$
D(\lambda,k)= \sum_{\mu}w_{m,\Delta}(\mu)
$$
where $m=a+1$ and $\Delta$ is described in Example 2. Moreover,
$D(\lambda,k)$ is equal to the dimension of the irreducible $sl_3$-module
with highest weight $(k-3b-2a)L_1-aL_3$.

\endproclaim

In particular, if $k=2a+3b$, the smallest level admissible for
$V_{\lambda}$,
then $D(\lambda,2a+3b)=$ dim $V_{-aL_3} = (a+1)(a+2)/2,$
see in [FH, 15.17] a formula for the dimension.

 \subhead Remark  \endsubhead
 Computation of $D(\lambda,k)$ was the starting point
of this work.

In the next sections we sketch a proof of Theorems 1 and 2.

\subhead
4. Formula for the Structure Constants
\endsubhead

Let $W^{\wedge}$ be the group of affine transformations of
the plane $\Cal P$
generated by reflections $s_1, s_2, s_3$, where $s_i$ is the reflection
at the line $y_i=0$ for $ i=1,2,$ and $s_3$ is the reflection at the line
$y_3=-3$.

Define another action of  $W^{\wedge}$ on $\Cal P$ by
$w\ast\lambda = w(\lambda - \alpha_3) + \alpha_3.$ Let
$\epsilon :  W^{\wedge} \to \{1,-1\}$ be the homomorphism taking reflections to
$-1$.

The structure constants of the Verlinde algebra $A(sl_3,k)$ are given
by the formula
$$
N_{\lambda,\mu}^{\mu + \nu} = \sum_{w \in W^{\wedge}} \epsilon(w)
\cdot d_{\lambda}(\nu + \mu - w \ast \mu),
\tag1
$$
if $\lambda,\mu,\mu+\nu$ are admissible at level $k$. This formula is an easy
combination of the definition of the
structure constants in [K,13.35] and
formula 12.31 in [FH].

\subhead
5. Proof of the Theorems
\endsubhead

Formula (1) holds if $\la, \mu, \mu+\nu$ are admissible at level $k$.
For fixed $\la$ and $\nu$, this means that $\mu$ belongs to the triangle
$$
\De_I = \{ \mu \in \Cal P \, | \, y_i(\mu) \geq 0,
y_j(\mu) \geq 0, y_{\ell}(\mu) \geq  - (\alpha_\ell, \nu) \},
$$
if $\nu$ is of type I, and to the triangle
$$
\De_{II} = \{ \mu \in \Cal P \, | \, y_i(\mu) \geq  - (\alpha_i, \nu),
y_j(\mu) \geq  - (\alpha_j, \nu), y_{\ell}(\mu) \geq 0 \},
$$
if $\nu$ is of type II.

We consider all terms of formula (1) as functions of $\mu \in \De_I$,
resp. of $\mu \in \De_{II}$.

Consider the following 13 elements of $W^{\wedge}$:
$$
S = \{ \, id, \,  s_a, \,
 s_bs_c, s_cs_b, s_bs_cs_b \, | \,  a=1,2,3,
 (b,c)=(1,2), (1,3), (2,3) \, \}.
$$

Rewrite (1) as

$$
N_{\lambda,\mu}^{\mu + \nu} = \sum_{w \in S} \epsilon(w)
\cdot d_{\lambda}(\nu + \mu - w \ast \mu) +
\sum_{w \in W^{\wedge}-S} \epsilon(w)
\cdot d_{\lambda}(\nu + \mu - w \ast \mu),
\tag2
$$

\proclaim{Lemma 1}
If $\la, \mu, \mu + \nu$ are admissible,
then all terms of the second sum in (2) are equal to zero.
\endproclaim

The lemma is an easy corollary of admissibility.

To prove Theorems 1 and 2 we compute explicitly 13 functions
of $\mu$
of the first sum in (2).

The function corresponding to $w=id$ is the constant function
$d_{\la}(\nu)$.

 From now on we assume that $\nu$ is of type I
and describe the remaining 12 functions.
 Type II is considered
similarly.

\proclaim {Lemma 2} The function
$d_{\la}(\nu + \mu - s_{\ell} \ast \mu)$ as a function of $\mu\in\De_I$

is equal to $d_{\la}(\nu)$, if $y_{\ell}(\mu) <
  z_{\ell,\la} -  (\alpha_\ell, \nu)$,

is equal to $0$, if
$y_{\ell}(\mu) \geq   z_{\ell,\la} + d_{\la}(\nu) - (\alpha_\ell, \nu) - 1$,

is equal to $ z_{\ell,\la}
+ d_{\la}(\nu) -  (\alpha_\ell, \nu) - y_{\ell}(\mu) - 1 $
otherwise. \newline
For $a=i,j$, the function $d_{\la}(\nu + \mu - s_{a} \ast \mu)$ as
a function of
$\mu\in\De_I$

is equal to $d_{\la}(\nu)$, if $y_a < z_{a,\la}$,

is equal to $0$, if $y_a \geq z_{a,\la} + d_{\la}(\nu) - 1$,

is equal to $ z_{a,\la} + d_{\la}(\nu) - y_{a}(\mu) - 1 $ otherwise.

\endproclaim

\proclaim {Lemma 3}
The function
$d_{\la}(\nu + \mu - (s_is_js_i) \ast \mu)$ as a function of $\mu\in\De_I$
is equal to $d_{\la}(\nu - (y_i(\mu)+ y_j(\mu)+2)\al_{\ell} )$.

The function
$d_{\la}(\nu + \mu - (s_is_j) \ast \mu)$ as a function of $\mu\in\De_I$
is equal
to
\newline
$d_{\la}(\nu - (y_i(\mu)+1)\al_{\ell} +  (y_j(\mu)+1)\al_j )$.

\endproclaim

Similar descriptions hold for the remaining functions of the set $S$.
Combining these descriptions we get Theorems 1 and 2.

\head References
\endhead

[F] G.Faltings, A proof of the Verlinde formula, J. Algebraic
Geometry, 3(1994) 347-374.

[FH] W.Fulton and J.Harris, Representation Theory, Springer-Verlag, 1991.

[GW] F.M.Goodman and H.Wenzl, Littlewood-Richardson Coefficients
for \newline
Hecke Algebras at Roots of Unity, Adv. in Math., 82(1990) 244-265.

[K] V.Kac, Infinite dimensional Lie algebras, Cambridge University Press,
Third Edition, 1990.

[V] E.Verlinde, Fusion Rules and Modular Transformations in 2D Conformal
Field Theory, Nucl. Phys.,B 300 (1988) 360-376.

[S] C.Sorger, La formule de Verlinde, Seminaire Bourbaki, 47eme annee, 1994-95,
n. 794, 1-23.

\end